\documentclass[aps,prd,eqsecnum,preprint,
amssymb,amsmath]{revtex4}
\usepackage{newtxtext,newtxmath}

\usepackage[T1]{fontenc}

\DeclareRobustCommand{\VAN}[3]{#2}
\let\VANthebibliography\thebibliography
\def\thebibliography{\DeclareRobustCommand{\VAN}[3]{##3}\VANthebibliography}

\usepackage{graphicx}	
\usepackage{amsmath}	
\usepackage{amssymb}	
\begin{document}




\def\be{\begin{equation}}
\def\ee{\end{equation}}
\def\bea{\begin{eqnarray}}
\def\eea{\end{eqnarray}}

\def\betap{\tilde\beta}
\def\del{\delta_{\rm PBH}^{\rm local}}
\def\Msun{M_\odot}
\def\fPBH{f_{\rm PBH}}
\def\OM{\Omega_M}
\def\OB{\Omega_B}
\def\ODM{\Omega_{DM}}

\newcommand{\dd}{\mathrm{d}} 
\newcommand{\Mpl}{\bar M_{\rm P}} 
\newcommand{\mpl}{m_\mathrm{pl}} 



\title{Primordial black holes from the QCD epoch: Linking dark matter, baryogenesis and anthropic selection}

\author{Bernard Carr$^{a,b}$, Sebastien Clesse$^{c}$, Juan Garc\'ia-Bellido$^{d}$\ }\email[]{B.J.Carr@qmul.ac.uk,  sebastien.clesse@ulb.ac.be,   juan.garciabellido@uam.es}
\affiliation{$^{a}$School of Physics and Astronomy, Queen Mary University of London, Mile End Road, London E1 4NS, UK}
\affiliation{$^{b}$Research Center for the Early Universe, University of Tokyo, Tokyo 113-0033, Japan}
\affiliation{$^{c}$Service de Physique Th\'eorique, Universit\'e Libre de Bruxelles (ULB), Boulevard du Triomphe, CP225, 1050 Brussels, Belgium}
\affiliation{$^{d}$Instituto de F\'isica Te\'orica UAM-CSIC, Universidad Auton\'oma de Madrid, Cantoblanco, 28049 Madrid, Spain}
\date{\today}


\begin{abstract}
If primordial black holes (PBHs) formed at the quark-hadron epoch, their mass must be close to  
 the Chandrasekhar limit, this also being the characteristic mass of stars. If they provide the dark matter (DM),
the collapse fraction must be of order the cosmological  baryon-to-photon ratio $\sim 10^{-9}$, which suggests a scenario in which a baryon asymmetry is produced efficiently in the outgoing shock around each PBH and then propagates to the rest of the Universe. We suggest that the 
 temperature increase in the shock provides the ingredients for
 hot spot electroweak baryogenesis.  This also explains why baryons and DM have comparable densities, the precise ratio depending on the size of the PBH relative to the cosmological horizon at formation. 
The observed value of the collapse fraction and baryon asymmetry
 depends on the amplitude of the curvature fluctuations which generate the PBHs and may be explained by an anthropic selection effect associated with the existence of galaxies. We propose a scenario in which the quantum fluctuations of a light stochastic spectator field during inflation
generate large curvature fluctuations in some regions, with 
the stochasticity of this field providing the basis for the required  selection. 
 Finally, we identify several observational predictions of our scenario that should be testable within the next few years. In particular,  the PBH mass function could extend to sufficiently high masses to explain the black hole coalescences observed by LIGO/Virgo.\\
\\
\end{abstract}

\maketitle


\section{Introduction}

Primordial black holes (PBHs) have been a focus of interest for nearly 50 years {\citep{1967SvA....10..602Z,Hawking:1971ei,Carr:1974nx}.
One reason for this is that only PBHs could be small enough for Hawking radiation to be important \citep{Hawking:1974rv}, those smaller than about $10^{15}$g  
having evaporated by now with many interesting cosmological consequences \citep{Carr:2009jm}. 
In the last few years, however, attention has shifted to PBHs larger than $10^{15}$g
because of the possibility that they can provide the DM. This idea goes  back to the earliest days of PBH research \citep{1975Natur.253..251C} but has only recently become popular ~\cite{Carr:2016drx,Frampton:2015xza,Garcia-Bellido:2017fdg,Clesse:2017bsw}. 
There are various reasons for this: (1) the failure to find alternative DM candidates;  (2) the fact that black holes definitely exist, so that one does not need to invoke new physics; (3) the realisation that PBHs circumvent  the well-known big bang nucleosynthesis (BBN) constraint on the baryon density,
since they formed in the radiation-dominated era. 

There is also possible observational evidence for PBH DM from a variety of 
effects~\citep{Carr:2020xqk}.
In particular, it has  been proposed that the gravitational waves from binary black hole  mergers detected by LIGO/Virgo~\citep{PhysRevX.6.041015,Abbott:2016blz,Abbott:2016nmj,LIGOScientific:2018jsj} could be explained by PBHs~\citep{Bird:2016dcv,Clesse:2016vqa,Sasaki:2016jop}.
This may not  require all the DM to be in the form of PBHs \citep{Ali-Haimoud:2017rtz,Raidal:2018bbj,Sasaki:2018dmp} but PBHs 
could  have many interesting cosmological and astrophysical effects even if their density is small.
For example, they have  been invoked to explain the seeding of the supermassive black holes in galactic nuclei \citep{Bean:2002kx,Clesse:2015wea,Carr:2018rid} or even galaxies themselves \citep{1966RSPSA.290..177H,1972ApJ...177L..79R,1984MNRAS.206..801C}, the generation of large-scale structure through Poisson fluctuations \cite{Meszaros:1975ef,1977A&A....56..377C,1983ApJ...275..405F,1983ApJ...268....1C,Afshordi:2003zb}, the 
correlation of the source-subtracted soft X-ray and infrared backgrounds \citep{2013ApJ...769...68C,Kashlinsky:2016sdv,2018RvMP...90b5006K} and a host of problems associated
with dwarf galaxies \citep{Silk:2017yai}.
Indeed, it has recently  been claimed that PBHs could explain seven distinct cosmological conundra \citep{Carr:2019kxo}. 

These effects are associated with a wide range of PBH masses and can be used to place interesting constraints on the number of PBHs, even if they play none of these roles. 
The constraints are 
usefully expressed as limits on the fraction $f(M)$ of the DM in PBHs of mass $M$ and have recently been reviewed 
in Ref.~\cite{Carr:2020gox}. There are only a few mass windows where PBHs could provide all the DM ($f = 1$): the intermediate mass 
range ($1-100\,M_{\odot}$), 
the lunar-mass range ($10^{20}-10^{24}$g), the asteroid-mass range ($10^{16}-10^{17}$g) and the Planck-mass range ($10^{-5}$g) for the relics of evaporation. However, these constraints usually assume that the PBHs have a monochromatic mass function (i.e. a spread of masses  $\Delta M \sim M$) and this is unlikely. 
Indeed, one  {\it requires} PBHs to have an extended mass function if they are to explain more than a single cosmological conundrum. A more careful analysis of the constraints is required in this situation \citep{Carr:2017jsz} and this may either help or hinder the PBH DM proposal~\citep{Green:2016xgy,Kuhnel:2017pwq}.

In this paper, we address a common criticism of the PBH scenario, that it requires fine-tuning of the initial collapse fraction. If we assume the PBHs all have a mass $M$ 
and form  
at  time $t$  in a radiation-dominated early Universe, then the collapse fraction at formation $\beta (M)$ is related to the current DM fraction $f(M)$ by $\beta (M) \sim  f(M) (t/t_{\rm eq})^{1/2} \sim 10^{-9}  f(M) (M/\Msun)^{1/2}$ where $t_{\rm eq}$ is the age of the Universe of matter-radiation equality and we assume that all PBHs have around the horizon mass, $M \sim 10^5(t/s)/\Msun$, at formation \citep{Carr:1975qj}.
Therefore $\beta(M)$ must be tuned to a tiny value to ensure $f(M) \sim 1$.  Were it much smaller, $f(M)$ would be negligible; were it much larger, PBHs would overdominate the cosmological density and there would not be enough baryons to make galaxies.   
There are many PBH formation mechanisms
but most of them depend on the generation of curvature perturbations, with the probability of collapse  being exponentially sensitive to their amplitude. Therefore the fine-tuning of the collapse fraction implies an even greater fine-tuning of the relevant cosmological parameters. 

Here we propose a specific scenario in which
 PBHs provide the DM and mostly form at the QCD epoch.  This idea  has a long history and was originally based on the idea that there may have been a 1st order phase transition at the QCD epoch \citep{Crawford:1982yz}. This is no longer plausible} but - for given spectrum of primordial fluctuations - one might expect PBH formation to be enhanced at this time because of the slight softening of the equation of state expected \cite{Jedamzik:1996mr,Jedamzik:1998hc,Cardall:1998ne,Schmid:1998mx,Sobrinho:2016fay,Iso:2017uuu}, a possibility recently revived by several authors~\citep{Byrnes:2018clq,PhysRevLett.122.101301}.
For a while the possibility that the DM could be PBHs formed at the QCD epoch seemed to be supported by microlensing observations \citep{Alcock:1996yv}. More  extensive data then appeared to exclude this possibility \citep{Tisserand:2006zx}. 
 {However, it has recently been claimed that microlensing constraints are less stringent 
 and may even allow 100\% of the DM  when one takes into account more realistic DM profiles~\cite{2017PhRvD..96d3020G,Calcino:2018mwh}, uncertainties in the detection efficiency~\citep{2011MNRAS.415.2744H,2015A&A...575A.107H} or the possibility that the PBHs are 
grouped into clusters~\citep{Garcia-Bellido:2017xvr}.  On the other hand, as reviewed 
by Ref.~\cite{Carr:2020gox}, there are now other limits -- for example, associated with supernova lensing \citep{Zumalacarregui:2017qqd} and accretion \citep{Inoue:2017csr} -- which further challenge the suggestion that QCD black holes could provide the DM.

Our own proposal, 
some aspects of which are described in more detail by Ref.~\cite{Garcia-Bellido:2019vlf}, combines two new  ideas: a PBH formation scenario involving an inflationary spectator field and an efficient phase of {\it hot-spot} electroweak baryogenesis induced by the out-of-equilibrium collapse to 
form PBHs.  In some regions of the Universe, a light spectator field, such as the QCD axion, can populate the slow-roll region of its potential, due to quantum stochastic fluctuations during inflation.   These regions undergo an extra expansion when the field dominates the density of the Universe, well after the end of inflation, which generates super-horizon curvature fluctuations.   These fluctuations collapse at the QCD epoch to form PBHs and, at the same locations, generate the baryons, which quickly propagate to the entire Universe, leading to  the observed matter-antimatter asymmetry before big-bang nucleosynthesis.   Our model therefore uses 
gravity to provide the local out-of-equilibrium and kinetic energy conditions for very efficient electroweak baryogenesis at the quark-hadron epoch.    
This naturally explains why the DM objects (i.e. PBHs formed at the QCD epoch) and visible stars have comparable masses, both of these being close to the Chandrasekhar limit.
It also explains why the collapse fraction $\beta$ at the QCD epoch is of order the  cosmological baryon-to-photon ratio $\eta$ and why the baryons and PBHs have similar densities. The latter coincidence is also unexplained in the standard model.

Although our model explains why $\beta \sim \eta$ if $f \sim 1$, it does not explain why these quantities have their actual observed values of around $10^{-9}$ (see Sec.~2.2).  However, it is well known that various constraints on the value of $\eta$ are required in order that galaxies can arise  \citep{1979Natur.278..605C}. These constraints can be regarded as ``anthropic''  selection effects, in which case our model implies that $\beta$ is also anthropically constrained.  
Although anthropic arguments are unpopular in some quarters, in recent years they have become more fashionable because the multiverse hypothesis allows them to be interpreted as a selection effect \citep{GarciaBellido:1994gb,
Linde:2002gj,2007unmu.book.....C}.
{Several other authors have also recently considered anthropic aspects of PBH formation \citep{2018PhRvD..98j3018A,Nakama:2018utx}.}
The most important  constraint concerns the closeness of the times of matter-radiation equality and photon decoupling. We argue that only Hubble patches where $\beta \sim \eta \sim 10^{-9}$ lead to the formation of galaxies. In patches with lower values, the universe remains radiation-dominated until long after decoupling and there are not enough baryons to make galaxies; in patches with much higher values, it becomes matter-dominated well before decoupling but most baryons are accreted by PBHs, which also prevents galaxy formation. In our scenario, the stochasticity of the spectator field during inflation naturally allows
such a selection effect but one could envisage other possibilities.

The plan of this paper is as follows. In Section 2 we discuss the mass and collapse fraction of PBHs forming at the QCD epoch, explaining why these are close to the Chandrasekhar mass and baryon-to-photon ratio, respectively. In Section 3 we propose a scenario in which PBH production at the QCD epoch naturally  generates a cosmological baryon asymmetry of order the PBH collapse fraction. In Section 4 we discuss how these two small but necessarily comparable numbers may be constrained by the anthropic requirement that galaxies can form, providing one has scenario in which the PBH collapse fraction varies in different regions. In Section 5 we describe a PBH formation scenario which allows this and involves a spectator field which drives a second  inflationary phase in some regions of the Universe; we analyse the form of the curvature fluctuations required to generate the PBHs  and show that the expected PBH mass function 
is consistent with current constraints. 
In Section 6 we draw some general conclusions.

\section{
PBH mass and collapse fraction}

PBHs could have formed at any time in the early Universe, with their initial mass taking any value above the Planck mass.
Invoking PBHs formed at the QCD epoch as the DM is attractive for three reasons: 
\begin{enumerate}
\item the quark-hadron transition may naturally enhance gravitational collapse due to the softening of the equation of state then; 
\item this  explains why the PBHs and visible stars both have masses comparable to the Chandrasehar mass; 
\item the PBH collapse fraction required to provide the DM is the baryon-to-photon ratio and this arises naturally if the PBHs are responsible for baryogenesis. 
\end{enumerate}
Point (i) has been emphasized by previous authors, most recently in Ref.~ \cite{Byrnes:2018clq}. As regards (ii), one expects a spectrum of masses, spanning the range $0.01-100 M_{\odot}$ in both cases. Indeed, the PBH mass function should
 have a series of peaks,
 corresponding to dips in the sound-speed at various epochs in the history of the Universe \citep{Carr:2019kxo}. We are mainly interested in the mass where most of the density resides  but an extended mass function has important implications for LIGO/Virgo observations since one expects the gravitational wave signal from binary PBH coalescences to peak at a larger value of $M$ than the density. In this section, we first focus on point (ii) and emphasize the link between the mass of the Hubble horizon at the QCD epoch and the Chandrasekhar mass. We then focus on point (iii) but leave 
 a detailed discussion of the scenario for explaining the coincidence between the collapse fraction and the baryon-to-photon ratio until later.
The key point is that the baryon asymmetry generated locally (i.e. around each black hole) is ${\cal O}(1)$ but this is reduced to $\beta$ after the diffusion of baryons to the rest of the universe.  This model also naturally explains another apparent fine-tuning -- the comparability of the PBH and baryonic densities. 

\subsection{Chandrasekhar and QCD epoch horizon mass coincidence}

The Chandrasekhar limit is the maximum mass of a white dwarf, this representing a balance between gravity  and the electron degeneracy pressure.  It can be shown to be~\citep{1931ApJ....74...81C} 
\begin{equation}\label{MCH}
M_{\rm Ch} = \frac{\omega}{\mu^{2}}\left(\frac{3\pi}{4}\right)^{1/2}\frac{M_{\rm P}^3}{m_{\rm p}^2}   \simeq
5.6\,\mu^{-2}\,M_\odot\,,
\end{equation}
where $M_{\rm P}$ is the Planck mass, $m_{\rm p}$ is the proton mass, $\omega \approx 2.018$ is a constant that appears in the solution of the Oppenheimer-Volkov (Lane-Emden) equation and $\mu$ is the number of electrons per nuclei (1 for hydrogen, 2 for helium). 
Stars more massive than $M_{\rm Ch}$ cannot avoid  gravitational collapse to a neutron star and ones somewhat larger than this collapse to a black hole. Therefore, this is  a lower limit on the mass of a black hole arising from stellar evolution. One can  also show that most hydrogen-burning main-sequence stars have a mass in the range $0.1$ to $10$ times $M_{\rm Ch}$ \citep{1979Natur.278..605C}. The lower limit comes from the nuclear ignition condition and the upper limit from the instability associated with radiation-pressure-dominated stars. 
In terms of fundamental units, one has $M_{\rm Ch} \sim \alpha_{\rm G}^{-3/2} m_{\rm p}$, where
\begin{equation}
 \alpha_{\rm G} \equiv G m_{\rm p}^2/ (\hbar c) = m_{\rm p}^2/M_{\rm P}^2 \approx 6 \times 10^{-39} 
\end{equation}
is the gravitational fine structure constant, so all stars have a mass within an order of magnitude of this.
More precisely,  $\alpha_{\rm G}^{-3/2} m_{\rm p} \approx 1.85 M_{\odot}$.

Let us now consider the mass of a black hole which forms from the gravitational collapse of a large curvature perturbation during the radiation era of the early universe. In this case,
some fraction $\gamma$ of the relativistic gas within the particle horizon collapses to form a PBH. The 
precise fraction is uncertain.
 The standard assumption used to be $\gamma \approx 0.2$ \citep{Carr:2009jm} but more recent numerical studies studies suggest $\gamma \approx 0.8$ (Musco, private communiction). During the radiation era, the density is $\rho_{\rm r} \approx 3/(32 \pi G t^2)$ and the Hubble horizon {(also particle horizon)} size is $d_{\rm H} \approx 2ct$, so the mass of a PBH forming at time $t$ is
\begin{equation}\label{MPBH}
M =  \frac{4\pi}{3} \gamma \rho_{\rm r}\,d_{\rm H}^3 \approx \frac{\gamma c^3t}{G} \, .
\end{equation}
We can express this in terms of the temperature, using the relation during radiation era~\citep{Weinberg:100595}
\begin{equation}
k_{\rm B} T = (16  g_* \pi^3  G t^2/45)^{-1/4} 
\approx 0.6 \, \alpha_{\rm G}^{-1/4} g_*^{-1/4} m_{\rm p} (t/t_{\rm p})^{-1/2} \, ,
\label{Trho}
\end{equation}
{where $k_{\rm B}$ is the Boltzmann constant,
$g_*$ is the number of relativistic degrees of freedom, 
$t_{\rm p} = \hbar/(m_{\rm p} c^2)  \sim 10^{-23}$ s is the proton timescale and henceforth we use units with $\hbar = c =1$.  At the QCD epoch, Eqs.~(\ref{MCH}), (\ref{MPBH}) and (\ref{Trho}) give 
\begin{equation}
\begin{aligned}
M &= \frac{\gamma\xi^2}{g_*^{1/2}}\left(\frac{45}{16 \pi^3}\right)^{1/2}
 \frac{M_{\rm P}^3}{m_{\rm p}^2} 
=  \left(\frac{15}{4}\right)^{1/2}\left[ \frac{\gamma \xi^2}{g_*^{1/2} \pi^2 \omega} \right] M_{\rm Ch} \\
&\approx 3.6 \left( \frac{\gamma}{0.8} \right)   \left( \frac{g_*}{10}\right)^{-1/2}   \left( \frac{\xi}{5}\right)^2 M_\odot\,,
\end{aligned}
\label{QCDmass}
\end{equation}
where $g_*$ is normalised appropriately, 
$\xi \equiv m_{\rm p} c^2/(k_{\rm B} T_{\rm QCD}) \approx 5$ is the ratio of the proton mass to the QCD transition temperature and we have put $\mu =1$ in the Chandrasekhar expression. 
The middle expression shows that PBHs naturally have around the Chandrasekhar mass 
if they form at the QCD epoch, the factor in square brackets being around $0.1$.

We stress that there are several important differences between a star and a PBH forming at the QCD epoch, despite ther similar masses. 
A region collapsing to a PBH has around the Hubble horizon size, $d_{\rm H} \sim (M/M_{\odot})$~km, at maximum expansion and does not collapse much before forming an event horizon, whereas stars have radii of order $10^6$~km and collapse by a factor of $10^6$.
Consequently the final spin of stellar black holes  is expected to be large, due to conservation of angular momentum, while that of PBH should be negligible {\citep{DeLuca:2019buf,Mirbabayi:2019uph}.} 

\subsection{PBH collapse fraction and baryon-to-photon ratio coincidence }

We denote the {\it total} fraction of the DM in PBHs (allowing for an extended mass function) by 
$f_{\rm tot}$ and the ratio of the DM and baryonic densities by $\chi$. The most recent Planck measurements~\citep{Aghanim:2018eyx} give the cold dark matter and baryon density parameters as $\Omega_{\rm c} h^2 = 0.12$ and $\Omega_{\rm b} h^2 = 0.022$, respectively, so 
\begin{equation}
 \chi \equiv \frac{\Omega_{\rm c}}{\Omega_{\rm b}} \simeq 5.5 \, . 
\end{equation}
The ratio of the PBH density to the baryonic density is then $\Omega_{\rm PBH}/ \Omega_{\rm b} = f_{\rm tot} \chi$ and this is constant after PBH formation (if we neglect PBH accretion) since they both scale as $a^{-3}$, where $a$ is the cosmic scale factor. 

The CMB density scales as $a^{-4}$, so  the  ratio of the PBH density to the CMB density scales as $a \propto t^{1/2}$ in the radiation era. 
The PBH collapse fraction is determined by {evaluating this ratio at} the PBH formation time, which we denote as $t_{\rm form}$.
To determine this, we need the relationship between the temperature and time. Before matter-radiation equality, this is given by Eq.~(\ref{Trho}).  If the collapse fraction at the PBH formation time   is $\beta$, this implies that at any subsequent epoch within the radiation era
\begin{equation}
\begin{aligned}
\frac{\rho_{\rm PBH}}{\rho_{\gamma}} &= f_{\rm tot}  \chi \,\frac{\rho_{\rm b}}{\rho_{\gamma}}
\approx 
0.4 \, f_{\rm tot} \chi \left( \frac{\eta m_{\rm p}}{k_{\rm B}T}\right) \\
&\approx 0.4 \left( \frac{16 \pi^3}{45} \right)^{1/4} f_{\rm tot} \chi \eta g_*^{1/4}  \alpha_{\rm G}^{1/4} \left(\frac {t}{t_{\rm p}} \right)^{1/2} \, ,
\end{aligned}
\label{pbhcmb}
\end{equation}
where $k_B$ is the Boltzmann constant and
\begin{equation}
\eta \equiv n_{\rm b}/ n_\gamma = 2.8 \times 10^{-8} \Omega_{\rm b} h^2 = 6.1 \times 10^{-10}~
\label{eta}
\end{equation}
denotes the ratio of the baryon and photon number densities.
We also use the relation
$\rho_{\gamma} \approx 2.7 n_{\gamma} k_{\rm B}T$~\citep{Weinberg:100595} and Eq.~(\ref{Trho}).
Since the ratio of  the PBH and photon densities can be written as $\beta (t/t_{\rm form})^{1/2}$, this implies 
\begin{equation}
\beta \approx 0.4 \left( \frac{16 \pi^3}{45} \right)^{1/4} f_{\rm tot}  \chi  \eta \, g_*^{1/4}  \, \alpha_{\rm G}^{1/4} \left( \frac{t_{\rm form}}{t_{\rm p}} \right)^{1/2} \, .
\label{beta1}
\end{equation}
We can also express this in terms of the mass of the PBH, Eq.~(\ref{MPBH}) implying that this is
\begin{equation}
M = \gamma \,\alpha_{\rm G}^{-1} m_{\rm p} (t_{\rm form}/t_{\rm p}) \, .
\end{equation}
Eqn (\ref{beta1}) then gives
\begin{equation}
\begin{aligned}
\beta &\approx 0.4 \left( \frac{16 \pi^3}{45} \right)^{1/4}  f_{\rm tot} \chi  g_*^{1/4}  \gamma^{-1/2} \eta  \alpha_{\rm G}^{3/4} \left( \frac{M}{m_{\rm p}} \right)^{1/2}  \\
&\approx   0.5 \,f_{\rm tot} [\chi \gamma^{-1/2} \eta g_*^{1/4}] \left( \frac{M}{M_{\odot}} \right)^{1/2} \, ,
\end{aligned}
\label{beta2}
\end{equation}
where the square-bracketed term is $10 \eta$ 
for $\chi \approx 5.5, \gamma = 0.8$
 and  $g_* \approx 10$.
The collapse fraction required for PBHs to provide the DM takes a very simple form if they are produced at the QCD transition.  This is because,  from Eqs.~(\ref{beta2}) and (\ref{QCDmass}), the collapse fraction then is
\begin{equation}
\beta \approx 0.4 f_{\rm tot}  \chi \eta  \xi
\approx 10 \, \eta \, ,
\label{beta3}
\end{equation}
where we have assumed  $f_{\rm tot} \approx 1$, $\xi \approx 5$  and
$\chi \approx 5.5$ at the last step.
This result is easily understood: since the temperature is  $0.2$ times the proton mass at the QCD epoch and the average photon energy is three times this, 
one has $\rho_{\rm b}/\rho_{\rm \gamma} \approx 2 \eta$ and $\beta$ is $5.5$ times this if the PBHs provide the dark matter.
Therefore no fine-tuning of the collapse fraction is required if there is some natural way of associating  
$\eta$ and
$\beta$. One might envisage three possibilities. 

\begin{itemize}
 
\item The photons may have been generated by the PBHs  (e.g. via accretion) in such a way that the photon-to-baryon ratio $S \equiv \eta^{-1}$ is  of order $\beta^{-1}$.  This is not impossible, although one needs most of the photons to have been generated by the BBN epoch.\\
\item The collapse fraction $\beta$ may have been determined by $\eta$ in some way. 
For example, since most antiprotons annihilate just before the QCD phase transition, leaving $1/\eta$ photons for each surviving proton, one just needs the surviving protons (or a least 80\% of them) rather than the photons to go into the PBHs. \\
\item The baryon asymmetry may have been generated by PBHs, so that $\eta$ is naturally driven to $\beta$ for PBH formation at the QCD transition.  This suggests a scenario in which there is efficient baryogenesis around the regions that collapse to PBHs (i.e. $\eta \sim 1$ locally) but with the baryons later propagating to the rest of the Universe (i.e.  $\eta \sim \beta$ after homogenization).
\end{itemize}

\noindent 
Since we have not found any convincing scenario for the first two possibilities,
we focus only on the third possibility in this paper. We next discuss a specific realisation of this proposal.

\section{Baryogenesis through  QCD black holes}
\label{baryogenesis}

In this section we argue that the observed baryon asymmetry may have been generated by PBHs in such a way that $\eta$ is naturally driven to $\beta$ for PBH formation at the QCD transition.  
The large curvature fluctuations, and subsequent gravitational collapse to PBHs upon horizon re-entry, would have provided the out-of-equilibrium conditions required for efficient baryogenesis
 within those regions,  leading to $\eta_{\rm local} \gtrsim 1$.  However, the plasma would have then homogenized before nucleosynthesis, distributing those baryons to the rest of the universe and
naturally explaining why $ \eta  \sim \beta $.   Our model is therefore different from previous attempts to relate PBHs and baryogenesis through evaporations~\cite{Dolgov:2000ht} or Affleck-Dine baryogenesis~\cite{2004PThPh.112..971M}.  It is described in more detail in Ref.~\cite{Garcia-Bellido:2019vlf}.

\subsection{Entropy production at PBH formation}

The collapse of a large curvature fluctuation to  a PBH at horizon re-entry is an extremely violent and far-from-equilibrium process. It is also responsible for a huge production of gravitational  entropy, which we now evaluate.
The entropy of the gas of relativistic particles within the horizon in the early universe can be written as~\citep{Weinberg:100595} 
\be
S_{\rm gas} = \frac{2\pi^2}{45}\,g_{*S}(T)\,T^3\ V_{\rm H}\,,
\ee
where $V_{\rm H}  \approx 32\pi t^3/3$ 
 is the horizon volume in the radiation era, $g_{*S}$ is the number of entropy states and we put $k_B= \hbar = c=1$ throughout this section. On the other hand, the gravitational entropy of a PBH formed from the gravitational collapse of this volume is~\citep{Hawking:1974sw}
\be
S_{\rm PBH} =  4\pi G M^2 \simeq 4\pi\gamma^2 \left( \frac{t}{t_{\rm P}} \right)^2\,,
\ee
where the PBH mass  is given by Eq.~(\ref{MPBH}) and $t_{\rm P}$ is the Planck time.
The ratio of these two quantities depends on the time of PBH formation, 
\be
\frac{S_{\rm PBH}}{S_{\rm gas}} = \gamma^2\left(\frac{405}{16\pi}\right)^{1/2}
\frac{g_*^{1/2}(T)}{g_{*S}(T)}\frac{M_P}{T} \ \simeq \ 0.5\times10^{20}\ 
\gamma^2\, \left(\frac{200\,{\rm MeV}}{T} \right)\,,
\ee
where we have used Eq.~(\ref{Trho}), so a PBH formed at the QCD epoch is responsible for an entropy increase which is huge compared with the entropy of the 
particles themselves.  
This provides an upper bound on the available entropy increase but the process described below requires significantly less entropy production and does not require all the PBH gravitational degrees of freedom.

\subsection{Energy production at PBH formation}

Since gravitational collapse is not 100\% efficient, a large fraction of the mass within the horizon is 
pushed away from the PBH by conservation of linear momentum. The expelled  particles are accelerated and acquire a kinetic energy equivalent to the difference in their potential energy  before and after the collapse. Since the radius of the PBH at formation is a factor $\gamma$ smaller than the horizon size, the total amount of kinetic energy released by the collapse into the remaining gas of 
relativistic and non-relativistic particles is 
\be\label{KE}
\Delta K \approx \left(\frac{1-\gamma}{\gamma}\right) \frac{c^5 t}{G} \, .
\ee
This is equivalent to  several solar masses of energy at the QCD epoch. The smaller the value of $\gamma$, the more the energy available.

We can now estimate the kinetic energy per particle generated by the collapse. 
Gravitational collapse is a very violent process, usually occurring
 in less than a Hubble time, so
 protons and neutrons remain out of equilibrium during the collapse and subsequent shock-wave formation, acquiring significantly more energy per nucleon than a thermal bath could provide.
Thus a large fraction have energy
\be
E_0 = \frac{\Delta K}{n\,V} \, ,
\ee
where $V$ is the volume of the shell around the collapsed region and $n$ is the number density of neutrons or protons at the time of collapse, this being determined by their thermal distribution just before then~\citep{Weinberg:100595}
\be
\begin{aligned}
n_{\rm n}(x) &\simeq n_{\rm p}(x) = 2\left(\frac{k_{\rm B} m_{\rm p}T}{2\pi \hbar^2}\right)^{3/2}\exp\left(-\frac{m_{\rm p} c^2}{k_{\rm B} T}\right)\\
&= 1.59\times10^{40} x^{-3/2}\,e^{-x}\,{\rm cm}^{-3}
\end{aligned}
\ee
with $x \equiv m_{\rm p} c^2/ (k_{\rm B} T)$. 
Not all protons acquire enough energy to induce sphaleron transitions but enough of them do to initiate baryon-number-violating processes. The kinetic energy each nucleon acquires from the out-of-equilibrium collapse of the PBH is 
\be
E_0 \simeq 140\,  \left( \frac{1- \gamma}{4 \gamma} \right) x^{-5/2}\,e^{x} \, {\rm GeV}\,,
\ee
where the term in brackets is $0.06$ for our favoured normalisation of  $\gamma = 0.8$. 
 
On the other hand, the particle density in
 the relativistic plasma surrounding the collapsed region
is huge then~\citep{Weinberg:100595},
\be
n_{\rm gas} = 1.64 \times10^{41} \ x^{-3} \,{\rm cm}^{-3} \,,
\ee
these being in thermal equilibirum. These particles constitute the target with which the  accelerated protons
collide, as in particle accelerators but at immensely larger densities. For example, at $T=120$ MeV, immediately after the QCD quark-hadron transition, one has protons with density $3\times10^{35}$ cm$^{-3}$ and kinetic energy $2$~TeV smashing into a wall of particles with density 
$3 \times10^{38}$ cm$^{-3}$.
At these energies and densities, the cross-section is of order a microbarn, so there is copious production of W-bosons within the Hubble time of 40 $\mu$s.

\subsection{Hot spot electroweak baryogenesis at the QCD epoch}

Since this is an extremely violent process,
the kinetic energies of protons and antiprotons is well above the plasma temperature and it produces 
 a high density gas of gauge bosons. Although this 
 occurs while the rest of the  Universe is well below the electroweak scale, the plasma is extremely hot locally and the rate of events is significantly enhanced compared
to the surrounding thermal state. The horizon that collapses is a hot fireball where high energy sphaleron transitions can take place very far from equilibrium, similar to the conditions
achieved with heavy ion collisions in high energy colliders but at much larger energies and densities. A somewhat related scenario was proposed in Ref.~\cite{2004PhRvL..92j1303A} in the context of low-scale reheating after inflation. We will show that these conditions are enough to generate the observed Baryon Asymmetry of the Universe (BAU). 

According toRef.~\cite{1991SvPhU..34..417S}, baryogenesis requires three ingredients: (1) baryon number violation; (2) C and CP violation; and  (3) out-of-equilibrium conditions to avoid any acquired asymmetry being washed out.
CP violation in the Standard Model (SM) is realized in the hadronic sector via the complex phases of the CKM (Cabibbo-Kobayashi-Maskawa) matrix~\citep{Zyla:2020zbs}.
The amount of CP violation is proportional to the Jarlskog determinant
 and given by~\citep{Cline:2006ts}
\be
\delta_{\rm CP}(T) = \frac{J}{T^{12}} \simeq \left(\frac{20.4\,{\rm GeV}}{T}\right)^{12}\,K \, ,
\label{delta}
\ee
with 
\bea
\begin{aligned}
&
J = (m_t^2-m_c^2)(m_t^2-m_u^2)(m_c^2-m_u^2) \\
& \times
(m_b^2-m_s^2)(m_b^2-m_d^2)(m_s^2-m_d^2) \, K  \, 
\end{aligned}
\eea
where $K = (3.06\pm0.2)\times10^{-5}$,
$m_t=172$ GeV, $m_b=4.5$ GeV, $m_c=1.27$ GeV,
$m_s=0.96$ GeV. 
Eq.~(\ref{delta}) shows that $\delta_{\rm CP}$ 
is extremely temperature-sensitive. 

At the classical level, the baryon and lepton symmetries are accidentally conserved in the SM.
However, the chiral anomaly implies that the currents are not conserved at the quantum level:
\be\label{NCS}
\partial_\mu j_B^\mu = \partial_\mu j_L^\mu = \frac{3\alpha_W}{8\pi}\,
F_{\mu\nu}\tilde F^{\mu\nu} \hspace{3mm} \Longrightarrow \hspace{3mm}
\Delta B = \Delta L = 3\Delta N_{\rm CS}\,,
\ee
where the Chern-Simmons number $N_{\rm CS}$ characterizes the different electroweak (EW) vacua and corresponds to the Higgs windings around its potential. Each winding generates a three unit baryon number jump.

The sphaleron rate $\Gamma_{\rm sph}$ describes the rate per unit 
time and volume 
at which long-wavelength configurations wrap around the SM false vacuum and make transitions from one Chern-Simmons number to the next. This induces the  baryon number violation given by  Eq.~(\ref{NCS}). The rate depends very strongly on temperature~\citep{Shaposhnikov:2000sja}:
\bea
\Gamma_{\rm sph}(T) \sim 
\begin{cases}
\alpha_W^4\,T^4 
& (T>T_c) \,, \\
\left(\frac{E_{\rm sph}}{T}\right)^3\,
m^4_W(T)\,e^{-\frac{E_{\rm sph}}{T}} 
& (T<T_c) \,,
\end{cases}
\eea
with $T_c\simeq 150$ GeV, $E_{\rm sph} \simeq 2\,m_W/\alpha_W$ and
$m_W^2(T) = \pi\alpha_W\,(v^2(T) + T^2)$.
Here $v(T) = v\,(1-T^2/12v^2)$ and 
$v=245$ GeV are the Higgs  vacuum expectation values at zero temperature. 

CP violation enters the dynamics through an effective operator~\citep{1999PhRvD..60l3504G}
\be
{\cal O} = \frac{3\alpha_W}{8\pi}\,\theta\,F_{\mu\nu}\tilde F^{\mu\nu}\,,
\ee
which induces an effective chemical potential for baryon production,
\be
\mu_{\rm eff} = \delta_{\rm CP}(T)\,\frac{\dd \theta}{\dd t} \, ,
\ee
with $\Delta\theta \sim \pi$ for each jump in $N_{\rm CS}$ and the CP-violation 
dimensionless parameter being given by Eq.~(\ref{delta}). 
The out-of-equilibrium evolution of the baryon number density $n_{\rm b}$ can be described by an {\it approximate} Boltzmann equation~\citep{1999PhRvD..60l3504G},
\be
\frac{\dd n_{\rm b}}{\dd t} + \Gamma_{\rm b} n_{\rm b} = \Gamma_{\rm sph}\,
\frac{\mu_{\rm eff}}{T_{\rm eff}}\,.
\label{boltz}
\ee
Here $\Gamma_{\rm b} = \frac{39}{2}\Gamma_{\rm sph}(T_{\rm th})/T^3_{\rm eff}$, where  $T_{\rm eff}$ is the effective temperature, giving the mean energy of the particles in the shock-wave, and $T_{\rm th}$ is the (lower) temperature of the thermal plasma surrunding the PBH. This term is
responsible for erasing the baryon density after baryon production, where but since the plasma surrounding the PBH collapse has a significantly lower temperature, the sphaleron transitions are quenched immediately. 
As long as the {right-hand side of Eq.~(\ref{boltz})} is large, the approximate solution  for $T_{\rm eff}\gg T_{\rm th}$ is \citep{1999PhRvD..60l3504G}
\be
n_{\rm b} = \int \dd t\,\Gamma_{\rm sph}(t)\,\frac{\mu_{\rm eff}(t)}{T_{\rm eff}}
\simeq \Gamma_{\rm sph}(T_{\rm eff})\,
\frac{\delta_{\rm CP}}{T_{\rm eff}}\,\Delta\theta\, .
\ee
This assumes that the sphaleron rate is dominant during the hot phase of the fireball expansion after the PBH collapse, at $T_{\rm eff} < 500$ GeV, while the CP violation is produced in the diffusion of those quarks and leptons through the surrounding thermal plasma towards a temperature around $T_{\rm th} = 70$ MeV. The entropy density at the end of this process is
 \be
 s=(2\pi^2/45)g_{*S}(T_{\rm th})T^3_{\rm th}
 \ee
 with $T_{\rm th}\sim 70\,{\rm MeV}$ and $g_{*S}(T_{\rm th})=10.75$. Therefore
\be
\eta = \frac{n_b}{n_\gamma} \simeq \frac{7n_b}{s} \ \simeq \ \frac{315}{2\pi^2g_{*S}}\frac{\Gamma_{\rm sph}(T_{\rm eff})}
{T_{\rm eff}\,T^3_{\rm th}}\,\delta_{\rm CP}\,\Delta\theta  \,,
\ee
which can be very large for $\delta_{\rm CP} \sim 1$.
Thus it is possible to generate a large BAU at the QCD scale, $\eta^{\rm local} \gg 1$, using only SM ingredients.   This means that the regions around the dense pockets that are collapsing to form PBHs  are saturated with baryons, so that the  local photon number can be highly suppressed.  The interplay between baryon-violating sphaleron transitions at the front of the shock-wave and the subsequent CP-violating processes that occur in the plasma surrounding the shell is reminiscent of the complicated processes that occur in the scenario of electroweak baryogenesis via first order phase transitions~\cite{1985PhLB..155...36K}. A detailed account of this complex phenomenon will require significantly more work than is reported here.

The final cosmological value of the BAU, $\eta\sim10^{-9}$, arises from the fact that the BAU is produced initially only in the hot spots around horizon domains that have collapsed to PBHs and then radiated away to the rest of the Universe. If the BAU occurs at the QCD scale (i.e. $t_{\rm QCD} \sim 10^{-5}$ s) and a fraction $\beta \sim 10^{-9}$ of all horizon volumes become PBHs, then the typical distance between PBH domains where the BAU has been generated is $d\sim \beta^{-1/3}\,d_{\rm H}(t_{\rm QCD}) \sim 10^{-2}$ light-seconds.  High-energy collisions in the expanding shock waves produce  jets  that uniformly distribute the localized BAU to the rest of the Universe. This  would then explain  the origin of the relation  $\eta \sim \beta$. 

In order for the protons and neutrons to redistribute the baryon number, they must travel through the plasma without a significant loss of energy and this assumption might be questioned. Protons are charged and will encounter electrons and photons with which they will interact, losing energy. However, neutrons only interact via their magnetic moments and these are suppressed at the relevant energies, so they can travel long distances before scattering. Since the plasma is hotter than 0.8 MeV before neutrino decoupling,
 the neutrons are redistributed
 throughout the universe and then converted into protons through weak interactions with the dense neutrino plasma. This regenerates the required proton abundance before BBN.
The weak interactions with the neutrino background on MeV scales are also insufficient to stop the neutrons and the QCD interactions (e.g. via pion exchange) are also suppressed below 20 MeV. 

 More precisely, the analytical expressions for the scattering cross-section of neutrons on electrons in a plasma at arbitrary energies give~\citep{PhysRevD.37.3441} 
\be
\frac{{\dd }E}{\dd t} = - \sigma_{\rm T}\left(\frac{31\pi^4}{252}\,G_{\rm M}^2\,T^6 -
\frac{7\pi^3}{30}\frac{m_{\rm e}^2}{m_{\rm n}^2}\,G_{\rm M}^2\,E^2\,T^4\right) \, ,
\ee
where $m_e$ and $m_n$ are the electron and neutron masses, $G_{\rm M} = -0.9$ is the Sachs magnetic form factor, and $\sigma_{\rm T}=8\pi\alpha^2/3m_{\rm e}^2$ is the Thomson cross-section, with $\alpha \equiv e^2/ (\hbar \, c) \sim 10^{-2}$ being the electric fine structure constant. For $E > 2\, k_{\rm B}T\,m_{\rm n}/m_{\rm e} \approx 4200 \, T$, the energy loss becomes a gain because of the minus sign. This happens at $E>300$ GeV for a $T=70$ MeV plasma, as applies in the period of PBH formation.  Ref.~\cite{PhysRevD.37.3441} assumes that the plasma is static at a fixed temperature, so they neglect the rapid expansion of the universe at PBH formation, this rapidly reducing the energy-loss term ($T^6 \propto t^{-3}$).  If one calculates the reaction rates, one finds that the neutrons travel from one PBH to the next before losing energy significantly. The situation is similar to primordial nucleosynthesis, where the nuclear reactions that would have generated all the light elements stop at lithium due to the cosmic expansion. 

Another relevant factor is that the conditions behind our QCD shock wave are similar to those in an evaporating black hole photosphere at the QCD temperature.  Heckler once claimed that an evaporating black hole forms an electron-positron photosphere (and later a QCD photosphere) due to the interactions of these particles among themselves \citep{Heckler:1995qq,Heckler:1997jv}.  However, Ref.~\cite{2008PhRvD..78f4043M} showed that they do not, the main reason being that the interactions are suppressed on account of the Landau-Pomeranchuk-Migdal effect. A similar suppression 
might apply in our context, although this is a complicated problem and requires further investigation. If it does, the extra suppression in the cross-section might allow both protons and neutrons to travel through the plasma over horizon distances before losing their energy, thus redistributing baryon number throughout the universe.  

The important feature of our scenario is that it  explains why the baryonic and DM densities are comparable. 
 Indeed, the value of $\chi$ should be  approximatively the ratio of the PBH mass and the ejected mass, $\chi \approx \gamma / (1-\gamma)$.  This is $5.5$, as observed, if $\gamma \approx 0.83$, a value compatible with simulations of PBH formation \citep{PhysRevD.100.123524}.  We note that if an overdensity of protons remains and eventually forms a shell around each PBH, then this would be accreted, resulting in no significant impact on the BBN constraints.
 
 \section{Fine-tuning of dark matter and baryon abundances}

The scenario proposed above explains why the PBH collapse fraction is of order the baryon-to-photon ratio and why the PBH density should be of the order of the baryon density. However, it does not explain the {\it actual} 
values of these quantities ($\beta \sim \eta \sim 10^{-9}$).
In this section, we suggest that an anthropic argument may be relevant 
because - independent of the nature of the DM - 
there are various constraints on the value of $\eta$ required in order that galaxies can form  \citep{1979Natur.278..605C}. In our model any constraint on $\eta$  implies an equivalent constraint on $\beta$ and hence on the amplitude of the curvature fluctuations on the PBH scale, so in this sense there would also be anthropic aspects to the collapse fraction.  Although this proposal is
 described as ``anthropic'',
it is really no more than a selection effect, providing one has a model in which different Hubble patches have different  fluctuations. In the next section we describe such a model but the considerations below apply more generally.

The most important  constraint concerns the closeness of the times of matter-radiation equality ($t_{\rm eq}$) and photon decoupling ($t_{\rm dec}$),  the first being only a little below the second. We will show that this corresponds to the combination of parameters $\eta (1+ \chi) $ (i.e. the entropy per DM particle) being roughly $\alpha^4$.
A larger value for this quantity (eg. a larger PBH density for a given value of $\eta$) 
would increase the gap between $t_{\rm eq}$ and $t_{\rm dec}$. A smaller value would imply $t_{\rm eq} \gg t_{\rm dec}$, which would also lead to problems.

We now discuss this argument in more detail, expanding on the analysis originally given
by Ref.~\cite{1979Natur.278..605C}. The photon number density, entropy density and energy density are
 \begin{equation}
\begin{aligned}
&
n_{\gamma} = \frac{30 \zeta (3) a_{\rm S} T^3}{\pi^4 k_{\rm B}}  \approx 0.37 \, \frac{a_{\rm S} T^3}{k_{\rm B}} \, ,
 \quad  s = \frac{4a_{\rm S}T^3}{3 k_{\rm B}} \approx 3.7\, n_{\gamma} \, , \\
&
\rho_\gamma = a_{\rm S} T^4 \approx 2.7 \, n_{\gamma} k_{\rm B} T \, ,
\end{aligned}
\end{equation}
respectively, where $a_{\rm S}$ is the black-body Stephan-Boltzmann constant.  
The photon temperature evolves as~\citep{Weinberg:100595}
\begin{equation}
k_{\rm B}T(z) = k_{\rm B}T_0 \,(1+z) = 2.5\times10^{-13}\,m_{\rm p}\,(1+z) \, ,
\end{equation}
so the redshift, density and temperature at matter-radiation equality are related by
\begin{equation}\label{MRE}
\rho_{\rm m} =  \left(\frac{1+z_{\rm eq}}{1+z} \right) \rho_{\rm r} = \frac{T_{\rm eq}}{T}\,  \rho_{\rm r}\,,
\end{equation}
where $\rho_{\rm m}$ is the total matter density (including any dark component). 
Since the ratio of the DM density (including any PBHs) to the baryon density is 
$\chi$, the total matter density is $1+\chi$ times the baryon density, so at matter-radiation equality
\begin{equation}
\rho_{\rm m} =  (1+\chi) n_{\rm b}  m_{\rm p}  =  \rho_{\rm r} \approx 2.7 \, n_\gamma  k_{\rm B} T_{\rm eq}  \, .
\end{equation}
 Then  Eq.~\eqref{eta} implies
\begin{equation}
 T_{\rm eq} \approx  0.37 \, (1+\chi) \eta \, k_{\rm B}^{-1} m_{\rm p}  \approx 2 \times 10^3 \, (1+\chi) \,  \rm{K} \, ,
\label{Teq}
\end{equation}
where $\eta$ is given by Eq.~\eqref{eta}.  
The Saha equation implies that the temperature at  decoupling is  less than the H-ionization energy by a factor of around $100$.
Since the ionization energy is $\alpha^2 m_{\rm e} c^2$, 
{where $m_{\rm e} $ is the electron mass,} this gives 
\begin{equation}
T_{\rm dec} \approx  0.01 \,\alpha^2 k_{\rm B}^{-1} m_{\rm e} \approx 0.1  \,\alpha^4 k_{\rm B}^{-1} m_{\rm p} \approx 3 \times 10^3\, \rm{K}  \, ,
\label{Tdec}
\end{equation}
where we have used the relation $m_{\rm e}/m_{\rm p} \approx 10 \,\alpha^2$, required for chemistry~\citep{1979Natur.278..605C}.
From Eq.~(\ref{Teq}), the (observed but unexplained) coincidence $T_{\rm eq} \sim  T_{\rm dec}$ corresponds to the condition
\begin{equation}
\eta \sim 0.1  \,(1+\chi)^{-1}\alpha^4 \sim 10^{-9} (1+\chi)^{-1}  \, ,
\label{etalpha}
\end{equation}
where we now use
order-of-magnitude relations. Alternatively, from Eqs.~(\ref{Teq}) and (\ref{Trho}), the time of matter-radiation equality is 
\begin{equation}
t_{\rm eq} \sim (1+ \chi )^{-2} \eta^{-2} \alpha_{\rm G}^{-1/2} t_{\rm p} \sim 10^{14} \, (1+ \chi )^{-2}  \rm{s} \, .
\label{teq1}
\end{equation}
After $t_{\rm eq}$, Eq.~(\ref{Trho}) is replaced by
\begin{equation}
T \sim T_{\rm eq} (t/t_{\rm eq})^{-2/3} \sim (1+ \chi )^{-1/3} \eta^{-1/3} \alpha_{\rm G}^{-1/3} m_{\rm p} (t_{\rm p}/ t )^{2/3}  \, ,
\end{equation}
so  from Eq.~(\ref{Tdec}) the time of decoupling (if later) is
\begin{equation}
t_{\rm dec} \sim  10 \, (1+ \chi )^{-1/2} \eta^{-1/2} \alpha_{\rm G}^{-1/2} \alpha^{-6} t_{\rm p} \sim 10^{13} \, (1+ \chi )^{-1/2} \rm{s} \, .
\end{equation}
The coincidence $t_{\rm eq} \sim t_{\rm dec}$ therefore corresponds to the condition (\ref{etalpha}), as expected.

In our scenario, $\chi$ is naturally ${\cal O}(1)$ because both the baryons and PBHs originate from the same collapsing curvature fluctuations, so the afore-mentioned coincidence 
depends 
on the condition $\eta \sim \beta \sim \alpha^4 \sim 10^{-9}$. Matter-radiation equality would  take place long  after decoupling for $\eta \sim \beta \ll 10^{-9}$, increasing the diffusion damping scale (the Silk mass) to well above that of a galaxy
and thereby suppressing the formation of such objects.
On the other hand,  it would occur long before decoupling for $\eta \sim \beta \gg10^{-9}$, so the diffusion damping scale is reduced, boosting the formation of structures with the size of dwarf galaxies and below.  In this case, PBHs would accrete most of the baryonic matter very quickly, 
 so there would not be enough baryons left over to make ordinary galaxies and the Universe would end up composed mainly of PBHs. 
If we drop the assumption that $\chi = {\cal O}(1)$,
which is specific to our baryogenesis scenario, then the above argument is more complicated and must be interpreted as an anthropic constraint on the number of PBHs for a given value of $\eta$. 
 Note that both $t_{\rm eq}$ and $t_{\rm dec}$ depend upon the combination $(1+\chi) \, \eta$, so it is the entropy per DM particle rather than the entropy per  baryon which is relevant.

There is another anthropic constraint on the value of $\eta (1+\chi)$.  By the  \cite{1961Natur.192..440D} anthropic argument, the current age of the Universe must be of order the main-sequence time of a star, which implies
\begin{equation}
t_0 \sim 100 \, \alpha_{\rm G}^{-1} t_p \sim 10^{17} \rm{s} \, .
\end{equation}
For life to arise, we require this time to exceed both $t_{\rm eq}$ and $t_{\rm dec}$. The first condition is more stringent and requires
\begin{equation}
\eta (1+\chi) \gtrsim \alpha_G^{1/4} \sim 10^{-10} \, .
\end{equation}
On the other hand, for the Universe to be radiation-dominated at BBN, we require
 \begin{equation}
\eta (1+\chi) \lesssim  \alpha_{\rm G}^{-1/4} (t_{\rm p}/t_{\rm NS})^{1/2} \sim 10^{-4} \, .
\label{bbn}
\end{equation}
where $t_{\rm NS}$ is the time of cosmological nucleosynthesis. Otherwise the Universe would be $100\%$ helium after after BBN and helium-burning stars might be too short-lived for life to arise around then. 
However, these arguments do not constrain $\eta (1+\chi)$ very tightly, which is why we have focussed on the condition \eqref{etalpha}. 

\section{Fine-tuning of curvature fluctuations}

Forming PBHs with the DM abundance requires curvature fluctuations larger than those expected 
 for slow-roll inflation in most single field scenarios~\cite{Garcia-Bellido:2017mdw,Ezquiaga:2017fvi}.  
 Invoking the softening of the equation of state during the QCD transition boosts
the formation of stellar-mass PBHs but does not alleviate the need for large curvature fluctuations.   
Furthermore, if the PBHs form from inhomogeneities, $\beta$ is exponentially sensitive to the amplitude of the power spectrum of curvature fluctuations, which accentuates the fine-tuning problem discussed above.
In this section, we propose a new PBH formation mechanism, involving a light stochastic spectator field, whose quantum fluctuations during inflation provide the key ingredient to resolving the fine-tuning.   Instead of being artificially fine-tuned, the mean value of this field within our observable Universe is explained by anthropic selection. 
We then propose identifying this scalar field with the QCD axion. 
This also allows
electroweak baryogenesis at the QCD epoch, as required for the model proposed in Sec.~\ref{baryogenesis}. 

\subsection{Basic idea}

The basic idea is that  quantum stochastic fluctuations in the spectator field during inflation lead it to
acquire different mean values in different current Hubble patches.
There are a huge number of these patches, so there necessarily exist some (eg. the one corresponding  to our
Universe) in which the spectator field on leaving the horizon has the value  required for subsequent quantum fluctuations to induce large curvature fluctuations over different horizon-sized regions at PBH formation (i.e. there are many such regions per current Hubble patch).
 More precisely,  the spectator field within these regions remains frozen during the radiation era until its potential energy starts to dominate the density of the Universe, well after inflation. At this point the field triggers a second inflationary phase (for at most a few e-folds) within these regions, whereas in the rest of the patch it quickly rolls down its potential without inflating.
This extra expansion generates local non-linear curvature fluctuations, which later re-enter  the horizon and collapse to form PBHs.
However, in the rest of the Universe the  curvature fluctuations are statistically Gaussian and  behave as expected in standard slow-roll inflation, unaffected by the spectator field.  

\subsection{The stochastic spectator during inflation}

We define three characteristics wave-numbers:  the scale of the observable Universe ($k_{\rm H_0} \simeq 2.3 \times 10^{-4}\, {\rm Mpc}^{-1}$), the CMB pivot scale ($k_{*}= 0.05\, {\rm Mpc}^{-1}$)
 and the PBH/QCD scale ($k_{\rm QCD} \simeq 10^6\, {\rm Mpc}^{-1}$).  There are about 22 e-folds of inflation between the observable Universe  and the PBHs  exiting the horizon and about 17 e-folds between  the CMB pivot and PBH scales doing so. During inflation we assume that the Hubble rate 
can be reconstructed from a truncated hierarchy of Hubble-flow slow-roll parameters,
  \be
  \epsilon_1 \equiv - \frac{\dd(\ln H) }{\dd N },  \hspace{0.5cm} \epsilon_{2} \equiv \frac{\dd (\ln \epsilon_1)}{ \dd N},  \hspace{0.5cm} \epsilon_{3} \equiv \frac{\dd (\ln |\epsilon_2|)}{ \dd N} \, ,
  \ee
 where $N$ is the number of e-folds since horizon exit of the CMB pivot scale.  
Then one has
 \begin{eqnarray}
 \epsilon_2 (N) &=& \epsilon_{2 *} \exp\left( {\int_0^N \epsilon_3(N')  \dd N'} \right)~, \\
 \epsilon_1(N) &=& \epsilon_{1* } \exp \left({\int_0^N \epsilon_2(N') \dd N' } \right)~, \\
  H(N) &=& H_{*} \exp \left({- \int_0^N \epsilon_1(N') \dd N' }\right)~.
 \end{eqnarray}
 Assuming slow-roll inflation, the scalar power spectrum amplitude $A_{\rm s} $ and the 
 spectral index $n_{\rm s} $ measured by Planck~\citep{Akrami:2018odb} are given to first order in the Hubble-flow parameters by
 \be
\begin{aligned}
&
 A_{\rm s} = 2.1 \times 10^{-9}  \simeq \frac{H_*^2}{8 \pi \epsilon_{1*}{\Mpl}^2 }~, \\
&
n_{\rm s} = 0.9649 \pm 0.0042 \simeq 1- 2 \epsilon_{1*} - \epsilon_{2*}~,
\end{aligned}
 \ee
 where $\Mpl=M_{\rm P}/\sqrt{8\pi}$ is the reduced Planck mass.
 
 Inflation is driven by a scalar field slowly rolling down its potential.   Three effective benchmark models are considered, 
 all agreeing with the amplitude and spectral index measurements: 
 \begin{itemize}
 \item Model 1: $ \epsilon_{1*} = 0.01, \epsilon_{2*} = \epsilon_{3*} = 2  \epsilon_{1*}=0.02, H_* =2.3 \times 10^{-5}  {\Mpl}$, as expected for a {\it quadratic} potential $V(\phi) \propto \phi^2$.\\
 \item Model 2: $ \epsilon_{1*} = 0.005, \epsilon_{2*} = \epsilon_{3*} = 4  \epsilon_{1*}=0.02, H_* =1.6 \times 10^{-5} {\Mpl}$, as expected for a {\it linear } potential $V(\phi) \propto \phi$.\\
 \item Model 3:  $ \epsilon_{1*} \lesssim 10^{-3}$ , $\epsilon_{2*} = 0.04$, $\epsilon_{3*}= 0, H_* \lesssim 7.3 \times 10^{-6} {\Mpl}$, as expected for {\it small-field} or {\it plateau-like} potentials.  
 \end{itemize}
 As explained later, if the spectator field dominates the density of the Universe much above the GeV scale, a significant decrease of $H(N)$ during inflation is needed to avoid an overproduction of light PBHs (i.e. $\epsilon_1 \gtrsim 10^{-3}$), which excludes Model 3
 but also predicts a detectable tensor-to-scalar ratio, $r \simeq 16 \epsilon_{1*}\gtrsim 2 \times 10^{-2}$.  Model 1 is disfavored by the current limits on this ratio, so Model 2 is preferred  in this case and the others should be regarded as  two extreme possibilities. 
However, the requirement on the Hubble rate variation during inflation is relaxed if the spectator field dominates
at the GeV scale or below, as in the case of the QCD axion, and Model 3 then leads to a more generic PBH mass distribution that is only marginally impacted by the exact shape of the inflationary potential.
This extends our scenario to any inflationary model.   Our approach is  relatively simplistic 
and does not involve 
advanced numerical methods
but it is precise enough to understand the 
physical principles behind our mechanism of PBH formation.    

We assume that there exists a light spectator field, $\psi$, with a potential $V(\psi) $ whose  shape is discussed later, having a mass $m_{\psi}\ll H_{\rm inf}$, where $H_{\rm inf}$ is the Hubble rate during the inflationary era.  In a coarse-grained model, the field experiences  stochastic quantum fluctuations. Because the field is very light, during one e-fold of inflation these are of order 
 $\Delta \psi^{\rm stoch} \sim H / 2 \pi$  within each Hubble volume. If the field is light enough not to reach adiabatic equilibrium during inflation, then the variance of the fluctuations $\delta \psi$ monotonically increases like
\be
\langle \delta \psi ^2 \rangle \simeq \int_0^N \frac{H(N')^2  }{4 \pi^2} \dd N'  ~.
\ee 
The evolution of $ \langle \delta \psi ^2 \rangle $ for our three benchmark inflation models is represented in Figure~\ref{fig:dpsi}.  If $\epsilon_1$ were constant, one would have
\be 
\langle \delta \psi ^2 \rangle \simeq \frac{H_{\rm CMB}^2}{8 \pi^2 \epsilon_1} \left( 1 - {\mathrm e}^{-2 \epsilon_1 N} \right)~,
\ee
which grows linearly with the number of e-folds before reaching a plateau when $N\gtrsim 1/(2 \epsilon_1)$.  In a realistic scenario $\epsilon_1$ varies during inflation, but qualitatively the spectator field variance follows a similar behavior.
The stochastic dynamics of the spectator field during inflation is described by the Fokker-Planck equation 
and admits a Gaussian solution for the probability density distribution~\citep{Hardwick:2017fjo},
\be  \label{eq:Ppsi}
P(\psi, N) = \frac{1}{\sqrt{2 \pi \langle \delta \psi^2\rangle} } \exp \left[ -\frac{ (\psi - \langle \psi  \rangle)^2 }{2 \langle \delta \psi^2 \rangle} \right]~,
\ee
where $\langle \psi  \rangle $ is the mean value within the patch corresponding to our Universe.    
One infers that the probability density of having a local field variation $\Delta \psi (N,x) \equiv \psi (N,x) - \psi (N-1,x) $ during one expansion e-fold is
\be
P(\Delta \psi, N) =  \frac{1}{\sqrt{2 \pi (H(N)^2/ 4 \pi^2) }} \exp \left[ -\frac{ \Delta \psi^2 }{2 (H(N)^2 / 4 \pi^2)} \right]~.
\ee
We will use this expression in the following section to compute the PBH abundance today.

\begin{figure*}[h!]
\begin{center}
\includegraphics[scale=0.8]{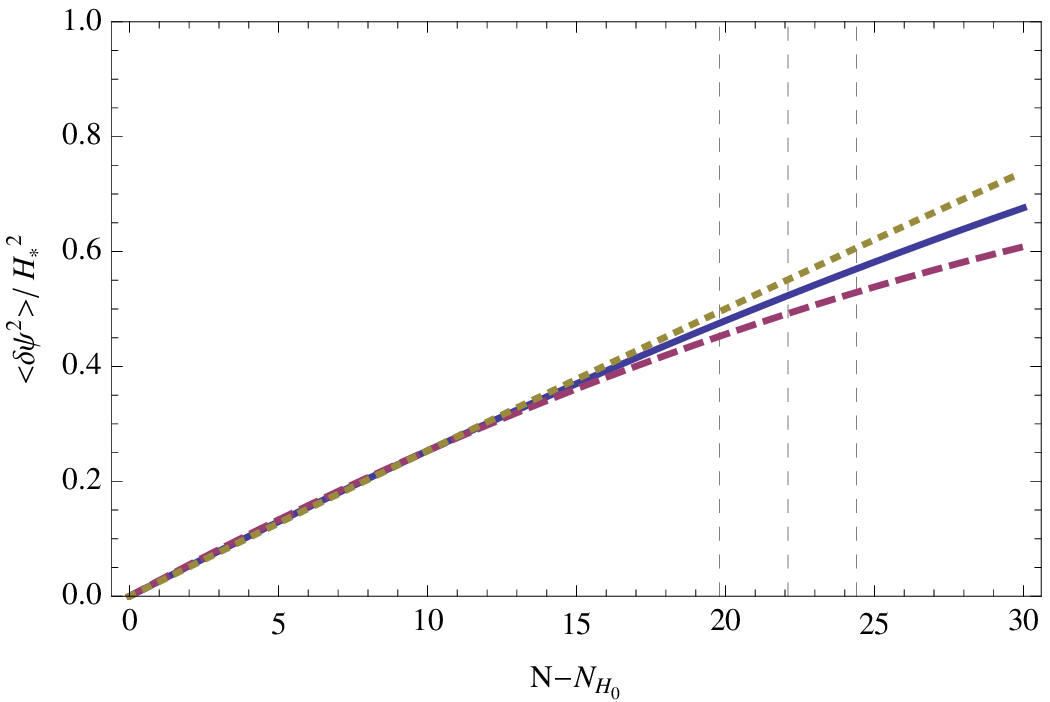} ~
\caption{Evolution of the spectator field variance as a function of the number of e-folds 
since the current Hubble scale, $k_{H_0}=2.3\times 10^{-4} {\rm Mpc}^{-1}$, exited the horizon
for models 1 (dashed red), 2 (solid blue) and  3 (dotted yellow).   The vertical dotted lines represent the number of e-folds for PBH masses of $0.01 M_\odot$, $1 M_\odot$ and $100 M_\odot$ (left to right). }
\label{fig:dpsi}
\end{center}

\begin{center}
\includegraphics[scale=0.7]{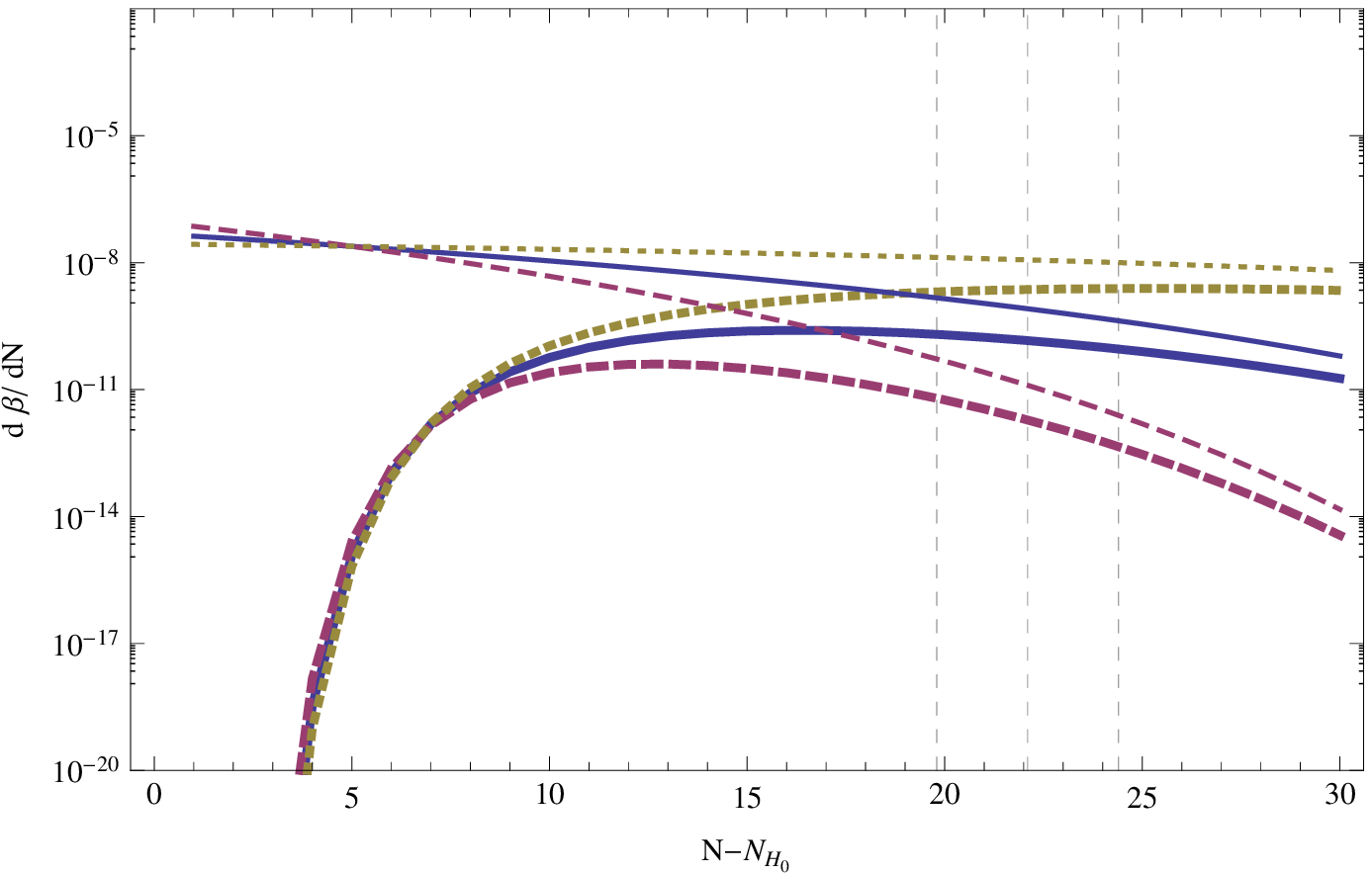} ~
\caption{
Collapse fraction of PBHs at formation, assuming a curvature threshold $\zeta_{\rm th} = 1.02$, as a function of  the number of e-folds since the {current Hubble} scale, $k_{H_0}=2.3\times 10^{-4} {\rm Mpc}^{-1}$, exited the horizon for models 1 (dashed red), 2 (solid blue) and  3 (dotted yellow), with $\epsilon_\psi = 0.5 H_{*}^2/ m_{\rm pl}^2$ and $\psi_{\rm cr} - \langle \psi \rangle = 10^{-4} H_{*}^2 $ (thin lines) or $\psi_{\rm cr} - \langle \psi \rangle = 0.5 H_{*}^2 $ (thick lines).  The vertical dotted lines are the number of e-folds for
PBH masses of $0.01 M_\odot$, $1 M_\odot$ and $100 M_\odot$ (left to right). }
\label{fig:spectrumreconstruction}
\end{center}
\end{figure*}

\subsection{The stochastic spectator field after inflation }

After inflation, the spectator field remains constant on super-Hubble scales as long as $m \ll H$ and $\rho \gg \rho_\psi \simeq V(\psi) $.   In the different patches, $\langle \psi  \rangle $ can take up to super-Planckian values if inflation lasts for a sufficient number of e-folds.   
Let us consider patches where  $\langle \psi  \rangle $ is close enough to field values for which the potential is sufficiently flat to induce slow-roll. For example,  the critical field value below which slow-roll is violated and inflation ends for a quadratic potential $V = m_\psi ^2 \psi^2$ is $\psi_{\rm cr} \simeq \sqrt 2 \Mpl$.  Then the spectator field starts to dominate the density of the Universe at some time during the radiation era and eventually induces in some regions a second (short) inflationary phase, generating large curvature fluctuations and leading to  PBH formation.  In other regions, the field quickly rolls down the potential without inflating.   In other patches where  $\langle \psi  \rangle $  is far from the slow-roll region, the stochastic field fluctuations are not able to produce a second inflationary phase in any region, so there is no PBH formation.   We focus here on the patches that have the value of  $\langle \psi  \rangle $ required for the subsequent field fluctuations to produce PBHs with an abundance compatible with the DM. 

We distinguish two possible behaviors, depending on the size of the slow-roll region generating $\mathcal O(1)$ curvature fluctuations in the spectator field potential, $\Delta \psi^{\rm sr}$, compared to the characteristic size of the quantum fluctuations during inflation, $\Delta \psi^{\rm stoch}$.

\textit{Case 1:  $\Delta \psi^{\rm sr} \gg \Delta \psi^{\rm stoch}$.}
From Eq.~(2.21) of Ref.~\citep{Hardwick:2017fjo},  the probability that the stochastic fluctuations lead the spectator field to acquire a local value $\psi > \psi_{\rm cr}$, the critical field value above which the slow-roll conditions are satisfied, is given by 
\be
P_1  = \int_{\psi > \psi_{\rm cr}} P(\psi, N) \dd \psi = \frac 1 2 {\rm erfc}  \left[  \frac{ \psi_{\rm cr}- \langle \psi \rangle }{\sqrt {2 \langle \delta \psi^2 \rangle} } \right]    \, . 
\ee
This provides a first condition for these regions to undergo an extra inflationary phase with $N_{\rm extra} \sim \mathcal O (1) $.   One can recognize a similar behavior when computing $\beta$ for Gaussian curvature fluctuations with a variance $\sigma$, viz. $\beta = {\rm erfc}(\zeta_{\rm tr} / \sqrt{2 \sigma^2}) $.     If $P_1$ were the only condition for PBH formation, since $\langle \delta \psi^2 \rangle $ is a growing function of time during inflation, the model would generally lead to an overproduction of light PBHs.   But PBH formation occurs only if a second condition is satisfied:  $\Delta \psi > \Delta \psi_{\rm tr}$ where $ \Delta \psi_{\rm tr}$ is the threshold fluctuation required to induce an extra e-folding $\delta N = \zeta_{\rm tr}$.  Indeed, in our coarse-grained picture, only these regions will experience  a curvature fluctuation (defined as the local curvature minus the mean curvature in the surrounding superhorizon region) leading to gravitational collapse when it re-enters inside the horizon.  This second condition has probability 
\be
P_2 = \int_{\Delta \psi > \Delta \psi_{\rm tr}} P(\Delta \psi, N) \dd \Delta\psi = \frac 1 2 {\rm erfc} \left[ \frac{\Delta \psi_{\rm tr} }{\sqrt 2 H_N / (2 \pi) } \right]~.
\ee
In this simplified pictured, PBH formation  occurs within one e-fold of expansion with probability
\be 
\begin{aligned}
P_{\rm PBH} &=
 \frac{ \dd \beta(t)}{\dd \ln M} = P_1 (N_t) \times P_2(N_t) \\
&= \frac 1 4 {\rm erfc}  \left[  \frac{ \psi_{\rm cr}- \langle \psi \rangle }{\sqrt {2 \langle \delta \psi^2 \rangle} } \right]   {\rm erfc} \left[ \frac{\Delta \psi_{\rm tr} }{\sqrt 2 H_N / (2 \pi) } \right]   \, , 
\end{aligned}
\label{PPBH}
\ee
where $N_t$ denotes the number of e-folds when the scale associated with  PBHs of mass $M $ exits the Hubble horizon.   The classes of plateau-like and large-field potentials correspond to this case.

\textit{Case 2:  $\Delta \psi^{\rm sr} \ll \Delta \psi^{\rm stoch}$.}   If the slow-roll region of the spectator field potential is tiny compared to its quantum fluctuations during inflation, 
 the probability of PBH formation is related to the probability that the field ends up in the slow-roll region, producing $\mathcal O (1)$ curvature fluctuation.  This assumes that the field distribution one e-fold earlier is given by Eq.~(\ref{eq:Ppsi}) with $N \rightarrow N-1$.   If one denotes by $\psi_{\rm min}$ and $\psi_{\rm max}$ the minimum and maximum field values 
in this region { (so that $\psi_{\rm max} =- \psi_{\rm min} = \Delta \psi^{\rm sr} $ for a symmetric potential), one obtains}
\be 
\begin{aligned}
&
P_{\rm PBH} = \int \dd \psi P(\psi, N-1) \\
&
 \times \frac{1}{2} \left[ {\rm erf} \left( \frac{\psi_{\rm max} - \psi } {\sqrt 2 H_N / (2 \pi)}\right) + 
{\rm erf} \left( \frac{\psi_{\rm min} +  \psi  }{\sqrt 2 H_N / (2 \pi)}\right) \right]~.
\end{aligned}
\ee
For a symmetric potential, in the limit $\Delta \psi^{\rm sr} \ll \Delta \psi^{\rm stoch}$, this gives
\be
P_{\rm PBH} = \int \dd \psi P(\psi, N-1)  \sqrt{\frac 2 \pi}  \frac{\Delta \psi^{\rm sr} }{H_{N} / (2 \pi)} \exp \left[ - \frac{\psi^2 }{2 H_N^2 /(4 \pi^2) }\right]~.
\ee
After integrating over the field distribution, one obtains
\be \label{PPBH2}
\begin{aligned}
&
P_{\rm PBH} =\sqrt{\frac 2 \pi}  \frac{\Delta \psi^{\rm sr}}{\sqrt{H_N^2/(4 \pi^2) + \langle \delta \psi^2 \rangle_{N-1}}  } \\
& \times  \exp \left[{- \frac{\langle \psi \rangle^2}{2 (H_{ N}^2 / 4 \pi^2 + \langle \delta \psi^2 \rangle_{N-1} )}} \right]~,  
\end{aligned}
\ee
which can be suppressed to any low value in patches where $\langle \psi \rangle  <  \sqrt{ \langle \delta \psi^2 \rangle} $.  }
 This mechanism allows PBH formation with $P_{\rm PBH} \sim 10^{-9}$ for 
\textit{small field}, \textit{double-well} or \textit{axionic} potentials.   {An important difference from the previous case is that, when $\langle \psi \rangle \ll \sqrt{H_N^2/4\pi^2 + \langle \delta \psi^2 \rangle_{N-1}} $, the PBH probability $P_{\rm PBH}$ becomes inversely proportional to $\sqrt{H_N^2/4\pi^2 + \langle \delta \psi^2 \rangle_{N-1}} $, instead of involving an erfc function.  As expected, it is roughly determined by the ratio of the width of the slow-roll region to the range explored by the field fluctuations.  If $H_N$ 
were
not drastically reduced during inflation, this mechanism would overproduce light PBHs.  Nevertheless, as in the first case, if  stochastic spectator domination does not occur too much  before PBH formation, as expected if the field is identified with  the QCD axion, this  naturally introduces a cut-off at small masses, so the model is viable.

\subsection{Short second phase of inflation}

Equations (\ref{PPBH}) and (\ref{PPBH2})  give  the probability that a region in our Universe will collapse to form a PBH when it re-enters 
the horizon during the radiation era.   
A large curvature fluctuation is generated by the short extra expansion induced when the spectator field slowly rolls towards the bottom of its potential.   One can use the  stochastic $\delta N$ formalism to link the local curvature fluctuation to this extra expansion, $\zeta(x) \approx \delta N (x) $, this  itself being due to a spectator field fluctuation $ |\Delta \psi |$ during inflation, which  remains  frozen until the field density dominates.    For simplicity, we  neglect the impact of the radiation density during the extra expansion and assume that the spectator field evolution respects the slow-roll conditions until $\psi$ reaches the critical value $\psi_{\rm cr}$ where the slow-roll parameter $ \epsilon_{\psi} \equiv \Mpl^2 (V'/V)^2 \approx 1$.   {
If $\psi_{\rm m} \equiv \rm max  (min) (\psi_{\rm cr}, \psi - \Delta \psi ) $ is the maximum (minimum) between the slow-roll region and the field value on immediate super-Hubble scales, the curvature perturbation  for an increasing (decreasing) potential is 
\be
\zeta(x) = \Delta N (x) =  \frac{1}{\Mpl^2} \int_{\psi_{\rm m}} ^{\psi} \frac{V(\chi)}{V'(\chi) }  \, \dd \chi  \, .
\ee}
Determining the distribution of curvature fluctuations more accurately  would require numerically  implementing the stochastic $\delta N$ formalism and solving the exact field and expansion dynamics for a large number of field trajectories and a given potential.  This is left for a future work.

One can get an approximation for $\zeta(x)$ by assuming that $\epsilon_\psi $ remains constant and that  
 the extra inflation ends abruptly when the field reaches the value $\psi_{\rm cr}$, such that 
\be
\zeta(x) \sim \frac{{\rm min} (   |\psi - \psi_{\rm cr} |,  |\Delta \psi  | ) }{\Mpl \sqrt{\epsilon_\psi (\psi)}} ~.
\ee
The resulting collapse fraction of PBHs at formation is represented in Fig.~\ref{fig:spectrumreconstruction} for the three considered models, assuming a curvature threshold  $\zeta_{\rm th} = 1.02$.
Since $ |\Delta \psi  | \sim H_{\rm inf}$, the condition for the curvature fluctuation to exceed the threshold for PBH formation is  $\epsilon_{\psi} \sim ~ H_{\rm inf}^2/\Mpl^2$, so the spectator field potential must be very flat. 
Note  that slow-roll is violated at $\psi_{\rm cr} \simeq \sqrt 2 \Mpl $ for the simplest quadratic or quartic potentials, so the
stochastic field fluctuations, $ \Delta \psi   \sim H_{\rm inf} \lesssim 10^{-5} \Mpl$, are unable to drive the field in the flat region of the potential where $\epsilon_{\psi} $ would be low enough to induce a large curvature fluctuation.   Instead, one needs plateau-like or small-field (e.g. double-well axionic) potentials in this context.  
Note also that, since the stochastic field quantum fluctuations are Gaussian, curvature fluctuations much larger than $\mathcal O(1) $ (i.e. deriving from an extra inflationary phase lasting more  than a few e-folds) are exponentially suppressed.  {On cosmological scales, matter and baryon density fluctuations are still correlated with curvature fluctuations, as when PBHs form from peaks in the power spectrum of Gaussian curvature fluctuations~\citep{Young:2015kda}.   In regions of the Universe with large positive curvature fluctuations, inflation ends and PBHs form slightly later, which induces a small overabundance of PBHs and baryons in this region.  Small isocurvature modes on cosmological scales could be present, as a second order effect from the particular evolution of the stochastic spectator during inflation, which marginally impacts the value of $P_{\rm PBH}$.}  In the next section, a more concrete and refined calculation is performed for the particularly interesting case in which  the stochastic spectator is identified with the QCD axion.

\subsection{PBH mass distribution and comparison with astrophysical constraints}

A black hole is formed when the curvature fluctuation exceeds some threshold $\zeta_{\rm tr}$.   The exact value of $\zeta_{\rm tr}$ depends on the equation of state $w = P/\rho $ and has been computed in the spherically symmetric situation analytically by Ref.~\cite{2017PhRvD..96h3517H} and using numerical relativity by Ref.~\cite{Musco:2012au}.  As already mentioned, PBHs in the stellar-mass range form during the QCD cross-over because the sound-speed reduction lowers the curvature threshold~\citep{Jedamzik:1996mr,Cardall:1998ne}.   Recently, Ref.~\cite{Byrnes:2018clq} has computed the expected PBH mass function more accurately, based on the latest results of lattice QCD simulations.  We adopt their methodology here.
 If the entire Hubble volume at re-entry collapses to form a PBH, one typically gets a peak in the PBH density at around $2-3 M_\odot$. More 
realistically, the size  is fixed by the mass inside the fluctuation at turn-around. Analytic 
considerations show that the final PBH mass could be reduced to one fifth of the Hubble mass ($\gamma \simeq 0.2$), corresponding to a peak at around $0.5 M_\odot$, but numerical calculations suggest $\gamma = 0.8$. Nevertheless fractions between 0.1 and 1 are also plausible. 

For \textit{Case 1}, the PBH mass functions obtained in our three  models are shown in Fig.~\ref{fig:massdist}, on the assumption that the PBHs provide all the DM (i.e. $f_{\rm DM} = 1$).  In order to avoid an overproduction of light PBHs, it is essential that either  the Hubble rate varies sufficiently during inflation, which is why we require  $\epsilon_1 \gtrsim 10^{-3}$, or that the spectator field dominates the energy density of the Universe below the TeV scale.  
A second (lower) peak is expected within the range $10-30 \, M_\odot $, on account of the softening of the equation of state associated with the annihilation of pions, and this might explain the LIGO/Virgo black holes.   
For \textit{Case 2}, with field domination above the GeV scale, the mass function is identical to that for Case 1 and Model 3. 

\begin{figure*}
\begin{center}
\includegraphics[scale=0.91]{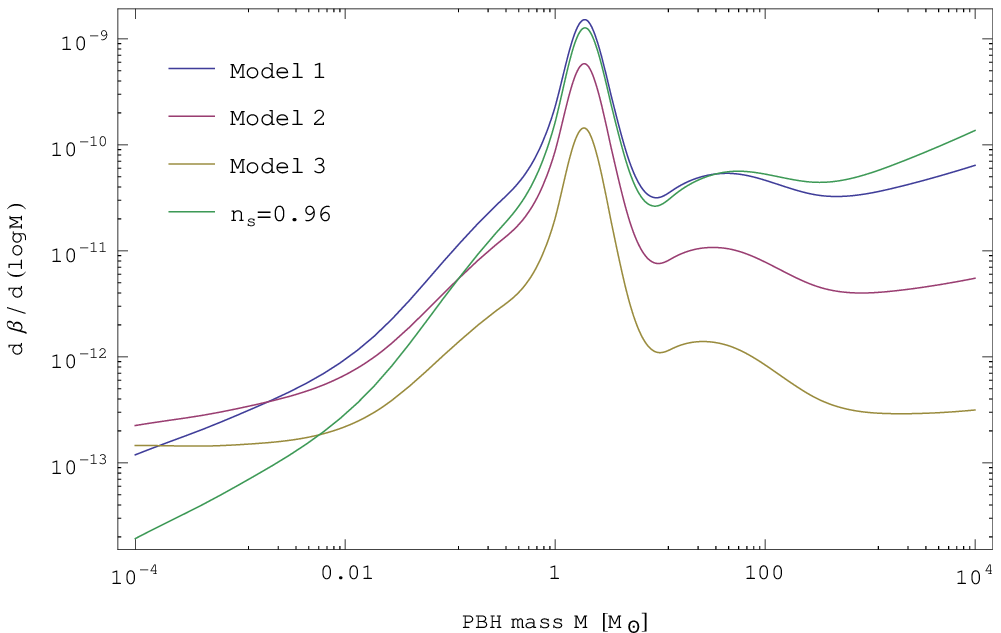} ~
\includegraphics[scale=0.91]{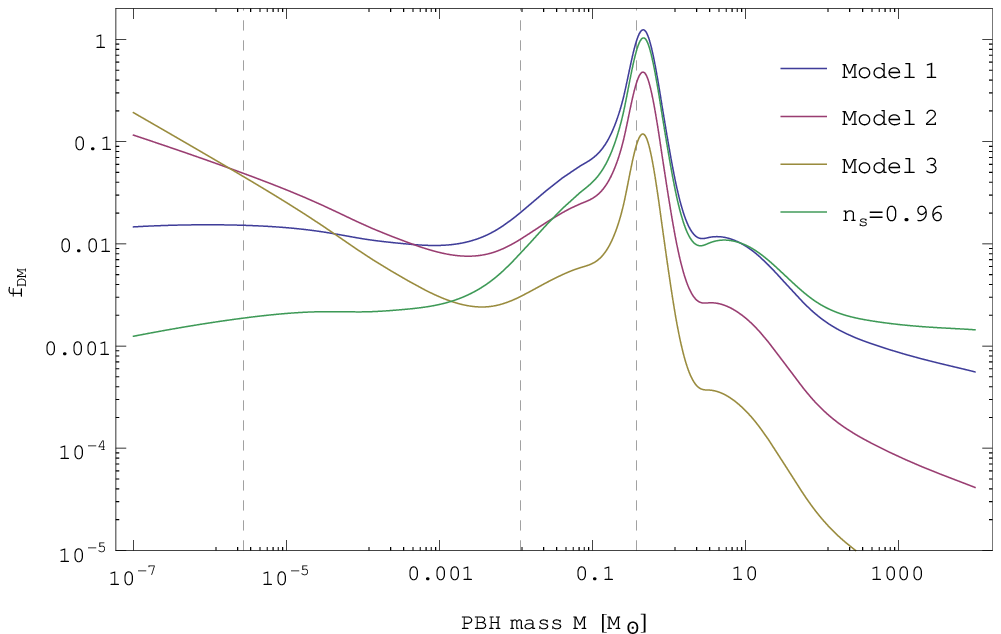}
\caption{
Top: Collapse fraction of PBHs at formation, assuming the equation of state parameter during the QCD cross-over given  by Ref.~\cite{Byrnes:2018clq}, 
for our three inflation models and an efficiency factor $\gamma = 0.8$. Here $\epsilon_\psi $ is chosen so that the PBHs provide all the DM and the spectator field potential corresponds to \textit{Case 1}.  
Bottom:  Corresponding PBH mass function, $f_{\rm DM} \equiv \dd \beta/ \dd \ln {M }$, the vertical lines representing the mass of PBHs formed at different temperatures. The mass function expected in the  model of Ref.~\cite{Byrnes:2018clq},
with $\gamma = 0.8$ and a nearly scale-invariant spectrum of curvature fluctuations with spectral index $n_{\rm s} = 0.96$, is also  shown.
}
\label{fig:massdist}
\end{center}
\end{figure*}

Let us now compare these predictions with the numerous observational constraints on the PBH mass function. 
These have recently been reviewed by Ref.~\cite{Carr:2020gox} but the most important ones in 
the mass range $1-100 M_\odot$ come from
dynamical heating of ultra-faint dwarf galaxies (UFDGs) and  
their star clusters \citep{Brandt:2016aco,Green:2016xgy,Li:2016utv,Koushiappas:2017chw},
anisotropies in the cosmic microwave background~\citep{Ricotti:2007au,Ali-Haimoud:2016mbv,Poulin:2017bwe}, non-observation of X-ray and radio sources towards the Galactic center \citep{Gaggero:2016dpq,Hektor:2018rul} or in the interstellar medium \citep{Inoue:2017csr} and the
non-disruption of wide stellar binaries in the galactic halo \citep{Quinn:2009zg,Monroy-Rodriguez:2014ula}.
There are also microlensing constraints but
these are weakened if the PBHs are clustered~\citep{Garcia-Bellido:2017xvr}.}

If the PBHs have a monochromatic mass distribution,  the general conclusion is that they could account for no more than $10\%$ of the DM in the mass range $1-100 M_\odot$.  The constraints are modified
 for an extended mass function \citep{
Carr:2017jsz} - in particular, for the lognormal mass distribution expected in many inflationary scenarios \citep{Dolgov:1992pu,2018JCAP...01..004B,Calcino:2018mwh} - 
but one can still exclude PBHs heavier than $10 M_\odot$ from providing most of the DM.  
Our scenario predicts that all the DM comprises PBHs but with the wide mass distribution shown in  Fig.~\ref{fig:massdist}.
This peaks around $0.5 \,\Msun$  and about $90 \%$ of the DM is in PBHs between $0.1$ and $2 \, M_\odot$.
Below $0.1 M_\odot$ and above $10 M_\odot$, PBHs contribute no more than a few percent of the DM, so one  evades all the above constraints. 
The peak is 
in the range probed by 
supernovae microlensing~\citep{Zumalacarregui:2017qqd} but some aspects of this constraint have been disputed ~\citep{Garcia-Bellido:2017imq}.   

An important probe of our model is the LIGO/Virgo events, which can 
be explained  as coming from the tail of the PBH distribution
~\citep{Clesse:2017bsw}.  The merging rates required are compatible with PBH
binaries  formed by capture in haloes at late times  if PBHs provide all the DM~\citep{Bird:2016dcv}.  The merging rate 
for those formed before matter-radiation equality does not allow more than 1\% of the DM  for a monochromatic mass function~\citep{Sasaki:2016jop} but 
the rate is suppressed for a wide-mass distribution because of binary  disruption by surrounding PBHs~\citep{Raidal:2018bbj}.   In our model, the merging rates of PBHs heavier than $10 M_\odot$ is additionally suppressed, since they contribute only a few percent of the DM.   
Searches for subsolar-mass black holes with gravitational wave experiments are also relevant.  The LIGO/Virgo observations already constrain the merging rates of subsolar equal-mass binaries~\citep{Abbott:2018oah,2018PhRvD..98j3024M} but again this applies only for a monochromatic distribution.   For a wide distribution, PBH mergers will only rarely have equal masses, so the rate will be spread out in the progenitor mass space and could be be impacted by early disruptions.      

Taking into account the uncertainties in the above constraints, it is still possible that all the DM comprises PBHs with an extended mass function of the kind predicted in our scenario.   Microlensing
surveys can clearly probe our model. For example, microlensing events in M31~\citep{CalchiNovati:2005cd,Niikura:2017zjd} are relevant and
recent observations of microlensing by dark lenses towards the Galactic bulge from the OGLE and Gaia surveys \citep{Wyrzykowski:2019jyg} favor a
population of black holes 
covering the mass gap between $2$ and $5 M_\odot $.  This is predicted in our scenario but  stellar black holes are not expected to form here.
Some of these issues are discussed in another recent paper  \citep{Carr:2019kxo}.

\section{Conclusions}

The model proposed in this paper
resolves and relates two of the most pressing problems of cosmology, the origin of the baryon asymmetry and the nature of DM.
Rather than relying on new particle physics interactions at high energy to generate the baryon asymmetry everywhere in the Universe simultaneously, our scenario suggests that it occurs only locally, in rare domains associated with PBH formation, and only later disperses to the rest of the universe. The BAU is generated by the violent process of PBH formation during the quark-hadron transition, this being triggered by the sudden drop in the radiation pressure in the presence of large-amplitude curvature fluctuations. Baryon number violation is driven by out-of-equilibrium sphaleron processes that are immediately quenched by the surrounding plasma in the expanding universe, preventing baryon wash-out, and the only CP-violation needed is provided by the CKM phases of the standard model. 

The key point is that the same small fraction of domains that act as hot spots for the efficient production of baryons is responsible for the 
 low value of the BAU and this also explains
why baryons and DM have similar densities today. 
The connection between the rareness of the domains, which is responsible for late matter domination, and the low baryon-to-photon ratio is a completely new way of approaching this problem and naturally links baryons and DM. 
 
Fine-tuning is still required  to explain why the baryon-to-photon ratio is of order $10^{-9}$ but this corresponds to a single anthropic selection argument associated with the formation of galaxies. The quantum fluctuations of a light stochastic spectator field during inflation, which are the basis for our anthropic selection argument, provide the rare super-horizon curvature fluctuations that are produced during a short transient phase well after inflation,  when the {spectator field dominates the density of the Universe.  These fluctuations collapse into solar-mass PBHs at the quark-hadron epoch
but an important feature of this scenario is that the
fluctuations are only local, and thus highly non-Gaussian, whereas  the curvature fluctuations in the rest of the Universe remain Gaussian and follow the predictions of standard slow-roll inflation. This avoids the need for an enhancement in the primordial power spectrum on some scale, which has long been considered unnatural and indeed one of the principal argument against PBHs. Note that  the existence of any light spectator field during inflation, as long as its density exceeds the QCD scale, inevitably leads to Hubble patches in which PBHs and baryons are formed with the observed relative abundances.

Several 
observable predictions of our model have been considered.  If the spectator field potential is of the plateau-type, with  $V^{1/4}$ being above the TeV scale, some variation of the Hubble rate during inflation is needed to avoid an overproduction of light PBHs.  This generates a tensor-to-scalar ratio $r \gtrsim 0.08 $, which would be in tension with {current} CMB observations {and easily detectable with upcoming ones.}   If $100 \, {\rm MeV} \lesssim  V^{1/4} \lesssim 1 \, {\rm TeV}$, this condition is relaxed and - for an inflation scale $H_{\rm inf} \lesssim 10^{-6} \Mpl $ - the generic PBH mass function is different from that expected for a nearly scale-invariant power-spectrum enhancement (see Fig.~\ref{fig:massdist}).  Another notable difference is the existence of a low-mass cut-off that depends on the energy scale at which the field starts to dominate the density of the Universe.  Stellar or quasar microlensing searches and subsolar PBH searches with gravitational wave interferometers are thus ideal for testing and distinguishing between the different scenarios.  

We have addressed various fine-tuning issues in this paper and - in concluding - we summarize the connection between them:

* The similarity of the DM (PBH) and baryon densities today ($\chi  \approx 6$) is unexplained in most models of PBH formation. This ratio is constant after baryogenesis and PBH formation but its actual value is unspecified and could be either much
larger or much smaller.  In our model, we expect $\chi \sim 1$ because the {\it local} baryon asymmetry generated around each PBH is ${\cal O}(1)$ {and we can predict its value rather precisely.}

* The usual criticism of the PBH DM proposal is that it requires fine-tuning of the PBH collapse fraction $\beta$. This needs to be tiny but not too tiny. Given the sensitivity of $\beta$ to the amplitude of the fluctuations, one would expect the current PBH density  to be either negligible or huge, leaving too few baryons to make galaxies in the latter case. However, in our proposal the collapse fraction is necessarily of order $\eta$  because the baryon asymmetry is {\it generated} by the PBHs. It is  ${\cal O}(1)$ locally around each domain that collapses to PBH, but reduced by the factor  $\beta$ after  the asymmetry has  diffused throughout the Universe. 

* As discussed in Section 4, there are long-standing tunings involving the baryon-to-photon ratio $\eta \sim 10^{-9}$.  It needs to be more than $\alpha_{\rm G}^{-1/4} \sim 10^{-10}$  to ensure the lifetime of  stars exceeds the time of matter-radiation equality but less than  $\alpha_{\rm G}^{-1/4}(t_{\rm p}/t_{\rm NS})^{1/2} \sim 10^{-4}$ to avoid all the Universe going into helium at cosmological nucleosynthesis. The comparability of the times of decoupling and matter-radiation equality requires the condition $\eta \sim \alpha^4$ and this 
might be interpreted anthropically.

* We have not addressed the other well-known fine-tuning problem: the comparability of the dark energy and DM densities ($\Omega_{\rm DE}/\Omega_{\rm CDM}  \approx 3$), this only applying at a particular epoch. This has recently been addressed by Ref.~\cite{Tzikas:2018wzd}, who  invoke a cosmological model in which the number of effective spatial dimensions is reduced from three to one at early times. However, this proposal is not compatible with our own  since the QCD epoch is long after the $1+1$ phase. 

In the present paper, we have presented a broad outline of our scenario and a more quantitative approach is required to derive accurate observational predictions.  Possible refinements would include a more accurate description of the stochastic dynamics of the spectator field during inflation, the computation of the exact spectator field dynamics after inflation using the $\delta N$ stochastic formalism, the details of the black hole collapse to extract a more accurate value of the efficiency factor $\gamma$, and a more accurate derivation of the DM to baryon ratio in our model of hot-spot electroweak baryogenesis model.  We should also consider in more detail some concrete realizations of our scenario, when the spectator field is embedded in a high-energy physics framework.   

Finally, there could also be interesting observational consequences related to the presence of extremely rare but large curvature fluctuations on cosmological scales.   For example, these could help explain the cold spot observed in the CMB or the existence of large and massive superclusters~\cite{2017ApJ...844...25B} that are statistically unexpected for Gaussian primordial fluctuations.   Gravitational backreactions from these rare but non-linear cosmic inhomogeneities might even mimic the dark energy~\citep{2004JCAP...11..010R}. Our scenario could thus provide a new framework to explain the same order of magnitude of the dark energy and DM densities today.

\section*{Acknowledgements}

We thank Mikhail Shaposhnikov for comments on electroweak baryogenesis, as well as Chris Byrnes, Alexander Dolgov and Karsten Jedamzik for useful remarks and discussions. We are also grateful to the referee for many suggestions for improving this paper. BC thanks the Research Center for the Early Universe (RESCEU) at University of Tokyo for hospitality received during this work. JGB acknowledges support from Research Project FPA2015-68048-03-3P [MINECO-FEDER] and the Centro de Excelencia Severo Ochoa Program SEV-2016-0597.  The work of SC is supported by the Belgian Fund for Research F.R.S.-FNRS.  

\bibliography{refs}








\end{document}